\documentclass[preprint,12pt]{elsarticle}

\usepackage[T1]{fontenc}
\usepackage[utf8]{inputenc}

\usepackage{lmodern}
\usepackage{microtype}
\usepackage{xcolor}

\usepackage{amsmath,amssymb,amsfonts}
\usepackage{amsthm}
\usepackage{mathtools}

\usepackage{tikz}
\usepackage{tikz-cd}
\usetikzlibrary{arrows.meta, calc, positioning}
\tikzcdset{arrow style=tikz, diagrams={>={latex}}}
\usepackage{geometry}
\usepackage{graphicx}

\usepackage{titlesec}
\makeatletter
\let\paragraph\elsparagraph
\@ifundefined{elssubparagraph}{}{\let\subparagraph\elssubparagraph}
\makeatother

\usepackage{empheq}

\usepackage{bm}
\usepackage{enumitem}
\usepackage{hyperref}
\usepackage[nameinlink,noabbrev]{cleveref}

\theoremstyle{plain}
\newtheorem{theorem}{Theorem}[section]

\theoremstyle{definition}

\newtheorem{remark}[theorem]{Remark}
\theoremstyle{note}

\journal{}

\begin{document}

\begin{frontmatter}

\title{A Homothetic Gauge Theory and the Regularization of the Point Charge}

\author{Fereidoun Sabetghadam}
\address{%
  \textit{Mechanical Engineering Faculty, Science and Research Branch, IAU, Tehran, Iran}\\
  \textit{Email: fsabet@srbiau.ac.ir}
}

\begin{abstract}
We introduce a Homothetic Hodge–de Rham (HHDR) theory that extends the de Rham complex and Hodge decomposition to homothetically dressed differential forms. The dressing, governed by a dilaton field and a Weyl weight $w$, defines the homothetic Hodge machinery. Imposing homothetic symmetry on physical laws yields scale–covariant interaction terms that arise canonically from the geometry and can be interpreted as penalty-type couplings in the language of differential equations. On this geometric foundation we construct a Homothetic Gauge Theory (HGT) for a general weight $w$ and then specialize to $w=1$ to formulate homothetic electromagnetism, obtaining homothetic Maxwell equations for a coupled system of the physical gauge field and a homothetic offset field. As a central application, we revisit the divergence of the self-energy of a point charge: modeling the charge as a boundary condition and choosing an appropriate dilaton profile we show that both the electric field and its total self-energy remain finite at the origin. The HHDR/HGT framework thus provides a mathematically controlled extension of gauge theory with potential implications for field theory, classical electrodynamics, and numerical penalty methods.
\end{abstract}


\end{frontmatter}

\section{Introduction}\label{section:Introduction}

The infinite self-energy of a point electric charge, arising from the $1/r^2$ behavior of the electric field, is one of the most persistent foundational problems in classical electrodynamics. It has motivated theoretical modifications over the entire last century~	\cite{rohrlich1990,frisch2008}, and has led to new physical models, as it often indicates that the theory is incomplete at short distances. Here we propose another one, a homothetic extension of the gauge theory.

Historically, attempts to regularize the point charge have followed several distinct paths. The earliest and most famous is the non-linear electrodynamics (NED) of Born and Infeld, which modifies Maxwell's Lagrangian to impose a maximum field strength, thereby rendering the self-energy finite~\cite{BornInfeld1934,SorokinReview2022}. Another approach involves introducing higher-derivative terms into the Lagrangian, as in the Bopp--Podolsky theory, which softens the short-distance behavior of the potential at the cost of introducing ghost states in the quantum theory~\cite{Bopp1940,podolsky1942}. More modern approaches often embed electromagnetism within a larger structure, such as string theory or Kaluza-Klein models, where couplings to additional scalar fields (dilatons) can alter the effective field strength~\cite{GibbonsMaeda1988,GarfinkleHorowitzStrominger1991}. Furthermore, coupling NED to general relativity also leads to regular black-hole solutions where spacetime singularities themselves are resolved~\cite{AyonBeatoGarcia1998,BronnikovReview2022}.

In this paper we develop a geometric framework for the point-charge problem. Working on a Weyl-integrable spacetime, we introduce a \emph{homothetic dilaton dressing} acting on differential forms. The dressing is realized as an affine map parameterized by a scalar dilaton field $\lambda(x)$ and an offset field $\alpha_d \in \Omega^d(M)$ (depending on the form degree $d$). We show that this affinity becomes linear upon embedding the usual complex $\Omega^\bullet(M)$ into the doubled complex $\bar{\Omega}^\bullet(M)=\Omega^\bullet(M)\oplus\Omega^\bullet(M)$, on which a group of homotheties $\mathcal{H}$ acts linearly.

This linearization is not merely a formal trick; it allows for the construction of a consistent mathematical and physical theory. We develop a \textit{homothetic Hodge--de Rham theory} on this doubled space, complete with a set of covariant block-operators ($\hat{d}$, $\hat{\delta}$, $\hat{\Delta}$) that respect the homothetic structure. The commutativity of these new operators with the homothetic transformation group allows us to establish a new \emph{homothetic Hodge decomposition}. While this complex is shown to be topologically isomorphic to the standard de Rham complex, its harmonic representatives are fundamentally different, as they are "dressed" by the dilaton field.

This new mathematical framework naturally gives rise to a \textit{homothetic $U(1)$ gauge theory}. By pulling back the standard Maxwell action into our doubled-field space, we derive the \textit{homothetic Maxwell equations} (HMEs). This resulting theory is a form of non-linear (homothetic) electrodynamics where new interaction terms, coupling the physical field to the offset field (which represents a new, dynamical gauge redundancy), emerge canonically from the geometry. Remarkably, we find that these geometrically-derived terms are identical in form to covariant penalty terms used in computational physics to enforce boundary conditions~\cite{Sabetghadam2015,Badri2020}.

The primary application of this formalism is the resolution of the point-charge problem. By modeling the charge not as a singular $\delta$-function source, but as a boundary condition on an infinitesimal sphere (a construction related to the theory of self-adjoint extensions), we solve the static, spherically symmetric HMEs. We demonstrate that for a suitable profile of the dilaton field near the origin---specifically, $\lambda(r) \sim -\alpha \ln r$ with $\alpha > 1/2$---the resulting electric field is regularized. This taming of the field's short-distance behavior leads directly to a finite total self-energy, thus resolving the classical divergence entirely within a self-contained, geometric framework.

This paper is organized as follows. In Sec.~\ref{section:2-First setup}, we introduce the Weyl-integrable geometry and postulate the homothetic dilaton dressing, which is parameterized by a general Weyl weight $w$, showing how it can be linearized in a doubled field space. In Sec.~\ref{section:4- Homothetic Hodge}, we develop the homothetic Hodge--de Rham theory, define the covariant block operators, and derive the homothetic Hodge decomposition. In Sec.~\ref{section:Homot_Elect}, we apply this formalism to construct homothetic electromagnetism, specializing to the case where the Weyl weight is unity ($w=1$). In Sec.~\ref{section: Homot-Cohomo}, we analyze the homothetic Lorenz gauge and show how the resulting field equations contain terms analogous to computational penalty terms. In Sec.~\ref{section:8}, we apply the full theory to the point-charge problem and explicitly demonstrate the regularization of the electric field and the finiteness of its self-energy. Finally, Sec.~\ref{conclusions} provides a summary of our results and discusses potential avenues for future research.

\section{Weyl integrable Spacetime and Homothetic Dilaton Dressing}\label{section:2-First setup}

\subsection{Classical Weyl-integrable geometry}\label{subsec:2.1}
Consider a Weyl-integrable spacetime $(M,[\tilde{\eta}],\phi)$, where $\phi:=d\lambda$ is a globally defined closed $1$-form and $[\tilde{\eta}]$ is the conformal class of Lorentzian metrics with representative $\tilde{\eta}=e^{-2\lambda}\eta$ (so that in the Einstein gauge $\lambda=0$ one has $\tilde{\eta}=\eta$, the Minkowski metric). Let $\nabla$ be the (Weyl-compatible) connection obeying
\[
\nabla_\tau \tilde{\eta}_{\mu\nu}=-2\,\phi_\tau\,\tilde{\eta}_{\mu\nu}.
\]
Then, $\nabla$ is invariant under the following gauge transformation \cite{Scholz2011,Condeescu2025}
\begin{equation}\label{eq:Weyl_gauge transf}
\tilde{\eta} \mapsto e^{-2\sigma}\tilde{\eta}, \qquad \phi \mapsto \phi+d\sigma.
\end{equation}
Along with the gauge freedom \eqref{eq:Weyl_gauge transf}, one usually considers the transformation
\begin{equation}\label{eq:Transf_Classic}
\alpha \mapsto e^{-w\sigma}\alpha,
\end{equation}
for a differential $p$-form $\alpha \in \Omega^p(M)$ \cite{KarananasMonin2016}, in which $w$ is called the \emph{Weyl weight}. In physics, it usually determines such that a physical theory be  invariant under the gauge transformation \eqref{eq:Weyl_gauge transf}. In the present work, we are going to extend this linear transformation.  
 
\subsection{Homothetic dilaton dressing}\label{subsection:2.2}

To achieve a non-singular field theory, based on the Weyl--integrable structure above, we postulate a \emph{homothetic dilaton dressing}
\begin{equation}\label{eq:HDD_1}
\boxed{
\tilde{\alpha}:=e^{-w\lambda(x)} \alpha + \left(1 - e^{-w\lambda(x)}\right) \alpha_d,}
\end{equation}
in which $\alpha$ is a $p$-form in Einstein gauge and $w$ is its Weyl weight. Also, $\alpha_d$ (the homothety center) is a distinguished scale-invariant differential $p$-form, which is called the \emph{offset field}. It introduces an additional degree of freedom and will be discussed later in $\S$ \ref{sec:offset}. 

Then, the gauge transformation \eqref{eq:Weyl_gauge transf}-\eqref{eq:Transf_Classic} changes to 
\begin{equation}\label{eq:New_gauge_trans}
\tilde{\eta} \mapsto e^{-2\sigma}\tilde{\eta}, \qquad \phi \mapsto \phi+d\sigma, \qquad \tilde{\alpha} \mapsto e^{-w(\lambda+\sigma)} \alpha + \left(1 - e^{-w(\lambda+\sigma)}\right) \alpha_d.
\end{equation}

\begin{remark} ~

\begin{enumerate}
\item The Weyl connection $\nabla$ is still invariant under the new gauge transformation.
\item For the special case $\alpha_d=0$, the conventional Weyl gauge freedom \eqref{eq:Weyl_gauge transf}-\eqref{eq:Transf_Classic} is retrieved for $(M,[\tilde{\eta}],\phi)$
\begin{equation}
\tilde{\eta} \mapsto e^{-2\sigma}\tilde{\eta}, \qquad \phi \mapsto \phi+d \sigma, \qquad \tilde{\alpha} \mapsto e^{-w\sigma}\tilde{\alpha}.
\end{equation}
However, for general $\alpha_d$, \eqref{eq:New_gauge_trans} introduces a homothety on the affine space $\alpha_d+\Omega^p(M)$, with the homothety center $\alpha_d$ and scale factor $e^{-w\lambda}$. Therefore, one can calls this theory the \emph{``homothetic extension of Weyl-integrable geometry''}.
\item On the above setting, the metric tensor is a 2-form with zero homothety center $\tilde{\eta}_d=0$, and the Weyl weight $w=2$.
\end{enumerate}

\end{remark}
However, the above dressing is not linear, but affine, and to restore linearity, we re-linearize it as follows. 

\subsection{Linearization}\label{subsection:2.3}

First we define an infinite-dimensional group of pointwise homothetic transformations acting fiber-wise:
\begin{equation}
\mathcal{H} := \left\{
\Lambda_w[\lambda] \mid \lambda \in C^\infty(M,\mathbb{R})
\right\}, \qquad
\Lambda_w[\lambda](x) :=
\begin{pmatrix}
e^{-w\lambda(x)} & 1 - e^{-w\lambda(x)} \\
0 & 1
\end{pmatrix},
\end{equation}
where the group operation is given by pointwise matrix multiplication:
\begin{align*}
(\Lambda_w[\lambda_1] \Lambda_w[\lambda_2])(x)
&= \Lambda_w[\lambda_1 + \lambda_2](x), \\
\Lambda_w[0](x) &= \mathbb{I}_2 =
\begin{pmatrix}
1 & 0 \\ 0 & 1
\end{pmatrix}.
\end{align*}
As one can see, $\mathcal{H}$ is isomorphic, as a Lie group, to the additive group $C^\infty(M,\mathbb{R})$ under pointwise addition.

\begin{remark}
Let $\mathfrak h := \mathrm{Lie}(\mathcal H)\cong C^\infty(M,\mathbb R)$ be the (abelian) Lie algebra of $\mathcal H$, with the pointwise bracket $[w\lambda_1,w\lambda_2]=0$.  
Consider the fiberwise representation $\rho:\mathcal H\to \Gamma(M,\mathrm{GL}_2(\mathbb R))$ given by
\[
\rho\big(\Lambda_w[\lambda]\big)(x)\;=\;
\begin{pmatrix}
e^{-w\lambda(x)} & 1-e^{-w\lambda(x)}\\[2pt]
0 & 1
\end{pmatrix}.
\]
Its differential at the identity is the Lie–algebra homomorphism
\[
d\rho_e:\mathfrak h\longrightarrow \Gamma\!\big(M,\mathfrak{gl}_2(\mathbb R)\big),\qquad
\lambda\longmapsto -w\,\lambda(\cdot)\,T,
\quad
T:=\begin{pmatrix}1&-1\\[2pt]0&0\end{pmatrix}.
\]
Equivalently,
\[
\left.\frac{d}{ds}\right|_{s=0}\Lambda_w[s\,\lambda](x)\;=\;-w\,\lambda(x)\,T,
\]
so $T$ (up to the sign convention coming from $\Lambda_w[\lambda]=e^{-w\lambda T}$) is the infinitesimal generator of the $\mathbb R$–action on each fiber. Consequently, the $\mathcal H$–exponential is computed pointwise via the matrix exponential:
\[
\Lambda_w[\lambda](x)\;=\;\exp\!\big(d\rho_e(\lambda)(x)\big)\;=\;\exp\!\big(-w\lambda(x)\,T\big)
\;=\;
\begin{pmatrix}
e^{-w\lambda(x)} & 1-e^{-w\lambda(x)}\\[2pt]
0 & 1
\end{pmatrix}.
\]
\end{remark}
A formal definition for the principal \texorpdfstring{$\mathcal{H}$}{H}-bundle is provided in \ref{app:principal_H_bundle}.\\

After definition of $\mathcal H$, now we define the embedding:
\begin{equation}
\label{eq:k-forms-embed}
\iota^\bullet : \Omega^\bullet(M) \hookrightarrow \bar{\Omega}^\bullet(M):=\Omega^\bullet(M) \oplus \Omega^\bullet(M),
\qquad
\alpha \mapsto \bar{\alpha}
= \begin{pmatrix}
\alpha \\ \alpha_d
\end{pmatrix}.
\end{equation}
Then, ${\mathcal H}$ has a well-defined pointwise action on $\bar{\Omega}^\bullet(M)$, symbolically we write $\hat{\Omega}^\bullet:={\mathcal H}\cdot \bar{\Omega}^\bullet$, with the definition:
\begin{equation}
\label{eq:k-forms-embed1}
\hat{\alpha} := \Lambda_w \cdot \bar{\alpha}
= \begin{pmatrix}
\tilde{\alpha} \\
\alpha_d
\end{pmatrix}
:= \begin{pmatrix}
e^{-w\lambda(x)} \alpha + \left(1 - e^{-w\lambda(x)}\right) \alpha_d \\
\alpha_d
\end{pmatrix}.
\end{equation}
As one can see, the homothetic dilaton dressing is embedded in a linear structure. This linearization enables us to keep our next structures linear, the merit that finally leads us to a homothetic Hodge decomposition.

The linearization may also be achieved via shifting the variables as we have done before \cite{Sabetghadam2026}. However, the above embedding in a higher dimension space yields a better separation of the homothetic group from the space of differential forms, which is more suitable for the gauge theory.

\noindent
\textbf{Other related works.}

The dilaton dressing has a long, widespread history, appearing in various guises across theoretical physics. In effective field theories, dilaton dressing has been used to capture spontaneous breaking of conformal symmetry \cite{Chacko2013, Bersini2023}, while in duality and T-plurality constructions, it preserves symmetry structures under field redefinitions \cite{Sakatani2022, SakataniUehara2022}. The gravitational dressing program further incorporates the dilaton to consistently couple matter fields to background geometries \cite{Giddings2024, Ketov2001}. In geometrically motivated approaches, dressing techniques — including the Weyl-dilaton frameworks — have recently been formalized through dressing field methods within Cartan geometry \cite{Francois2019, Francois2025}. These applications differ in scope and purpose. In the present work, the homothetic dilaton dressing is introduced to preserve regularity, ultimately enabling a non-singular field-theoretic structure.

\section{A Homothetic Hodge Decomposition}\label{section:4- Homothetic Hodge}

The relations between different ranks of the above homothetically dilaton-dressed (HDD) fields is investigated here, after developing some operators we need.  
\subsection{Homothetic de Rham complex and Hodge theory}\label{sub-section:4.2}

Although homology, cohomology, and the exterior derivative are purely topological structures and therefore independent of the manifold's metric, the situation is different for the Hodge operators---such as the Hodge star, the codifferential, and related constructs---which explicitly depend on the metric. Moreover, the homothetic dilaton dressing (HDD) introduces a scale factor into the exterior algebra. Consequently, the key step in analyzing the HDD differential forms is to construct the appropriate covariant operators on this geometry. In this regard, we first define:
\begin{equation}\label{eq:Matrix-Operators}
\bar{d} := 
\begin{pmatrix}
d & 0 \\
0 & d
\end{pmatrix}, \quad
\bar{\delta} :=
\begin{pmatrix}
\delta & 0 \\
0 & \delta
\end{pmatrix}, \quad
\bar{\Delta} :=
\begin{pmatrix}
\Delta & 0 \\
0 & \Delta
\end{pmatrix},\qquad\
\bar{\star} := 
\begin{pmatrix}
\star & 0 \\
0 & \star
\end{pmatrix}, \quad
\end{equation}
where $d$ is the exterior derivative, $\delta$ is the codifferential, $\Delta = d\delta + \delta d$ is the Hodge–de Rham Laplacian, and $\star$ is the Hodge star operator. Moreover, we define the twisted operators
\begin{equation}\label{eq:Twisted-Operators}
\tilde{d}:=e^{-w\lambda} ~d ~e^{w\lambda}, \qquad \tilde{\delta}:=e^{-w\lambda} ~\delta ~e^{w\lambda}, \qquad \tilde{\Delta}:=e^{-w\lambda}~ \Delta ~e^{w\lambda}=\tilde{d}\tilde{\delta}+\tilde{\delta}\tilde{d}. 
\end{equation}
Then, we define the Weyl covariant counterparts of operators \eqref{eq:Matrix-Operators}:
\begin{equation}\label{eq:Weyl-cov-operators}
\hat{d} :=
\begin{pmatrix}
\tilde{d} & d-\tilde{d} \\
0 & d
\end{pmatrix}, \quad 
 \hat{\delta} :=
\begin{pmatrix}
\tilde{\delta} &  \delta -\tilde{\delta} \\
0 & \delta
\end{pmatrix}, \quad
 \hat{\Delta} := \hat{\delta} \hat{d} + \hat{d} \hat{\delta}=
 \begin{pmatrix}
\tilde{\Delta} &  \Delta -\tilde{\Delta} \\
0 & \Delta
\end{pmatrix},
 \quad
 \hat{\star}:=\bar{\star}.
\end{equation}

\begin{figure}[t]
  \centering
  \begin{tikzcd}[column sep=2em, row sep=2em, arrows={line width=0.7pt}]
    \cdots \arrow[r] &
    \bar{\Omega}^{p-1}(M) \arrow[r, "\hat d"] &
    \bar{\Omega}^{p}(M) \arrow[r, "\hat d"] &
    \bar{\Omega}^{p+1}(M) \arrow[r] &
    \cdots \\
    \cdots \arrow[r] &
    \bar{\Omega}^{p-1}(M) \arrow[r, "\bar d"] \arrow[u, "\mathcal H"] &
    \bar{\Omega}^{p}(M) \arrow[r, "\bar d"] \arrow[u, "\mathcal H"] &
    \bar{\Omega}^{p+1}(M) \arrow[r] \arrow[u, "\mathcal H"] &
    \cdots \\
    \cdots \arrow[r] &
    \Omega^{p-1}(M) \arrow[r, "d"] \arrow[u, hook, "\iota"] &
    \Omega^{p}(M) \arrow[r, "d"] \arrow[u, hook, "\iota"] &
    \Omega^{p+1}(M) \arrow[r] \arrow[u, hook, "\iota"] &
    \cdots
  \end{tikzcd}
  \caption{There is a chain isomorphism $\mathcal H\circ\iota^\bullet$ between the ordinary de Rham complex and the homothetically dilaton dressed complex.}
  \label{fig:triple_deRham_segment_tight_v2}
\end{figure}

The key common feature of these operators is their covariance with the homothetic transformations (verifiable directly):
\begin{equation}
\label{eq:commut-diff-forms}
\boxed{
\hat{d}\Lambda_w=\Lambda_w\bar{d}, \qquad \hat{\delta}\Lambda_w=\Lambda_w\bar{\delta}, \qquad \hat{\Delta}\Lambda_w=\Lambda_w\bar{\Delta}, \qquad \bar{\star}\Lambda_w=\Lambda_w\bar{\star},}
\end{equation}
and as the direct consequences of these features, one can verify that irrespective of the rank:
\begin{enumerate}
\item Both $\hat{d}$ and $\hat{\delta}$ are \emph{nilpotent}, that is, $\hat{d}^2=0$ and $\hat{\delta}^2=0$.
\item With the conformally scaled metric $\tilde{\eta}=e^{-2\lambda(x)}\eta$, $\hat{\delta}$ is the \emph{adjoint} of $\hat{d}$, that is~\footnote{It should be noted that without the conformal scaling $\eta\to\tilde{\eta}$, $\hat{\delta}$ is not the formal adjoint of $\hat{d}$. It may be seen as an evidence of internal consistency.}:
\begin{equation}
\langle \hat{d}\hat{\alpha},\hat{\beta}\rangle_{\tilde{\eta}}=\langle\hat{\alpha},\hat{\delta}\hat{\beta}\rangle_{\tilde{\eta}}
\end{equation}
\item As a consequence of adjointness of $\hat{d}$ and $\hat{\delta}$, one can say that $\hat{\Delta}$ is \emph{elliptic}.
\item The block Hodge star operator $\bar{\star}$ and $\Lambda_w$ commute without any further modification. 
\end{enumerate}

Now, we define the homothetic de Rham complex $\big(\hat{\Omega}^\bullet(M),\hat d\big)$ with
\begin{equation}\label{eq:Omeg_hat}
\hat{\Omega}^\bullet(M):=\operatorname{Im}\big(\mathcal H\circ \iota^\bullet\big)\subseteq\bar{\Omega}^\bullet(M), 
\end{equation}
where $\iota^\bullet$ is the embedding \eqref{eq:k-forms-embed}, and $\mathcal H$ acts fiberwise by the homothetic dressing. One can show that $\hat{\Omega}^\bullet$ is an affine subspace of $\bar{\Omega}^\bullet(M)$, and canonically isomorphic (as a graded vector space and as a chain complex) to the usual de Rham complex $(\bar{\Omega}^\bullet(M),d)$ (see Fig.~\ref{fig:triple_deRham_segment_tight_v2}.)

Moreover, we observe that the map $\mathcal H\circ\iota^\bullet:\big(\Omega^\bullet(M),d\big)\to\big(\hat{\Omega}^\bullet(M),\hat d\big)$ is a \emph{chain isomorphism}: using $\hat d\,\Lambda_w=\Lambda_w\,\bar d$ and $\bar d\circ\iota^\bullet=\iota^\bullet\circ d$ we obtain the commuting relation
\begin{equation}\label{eq:homothetic_chain_iso}
\hat d\circ\mathcal H\circ\iota^\bullet=\mathcal H\circ\iota^\bullet\circ d,
\end{equation}
which is precisely the vertical commutativity illustrated in Fig.~\ref{fig:triple_deRham_segment_tight_v2}. In particular,~\eqref{eq:homothetic_chain_iso} implies that the ordinary de~Rham cohomology and the homothetic (hatted) cohomology (defined in \eqref{homCoh}) are naturally isomorphic,
\[
H^k_{\mathrm{dR}}(M)\;\cong\;\hat H^k(M),
\qquad
[\alpha]\longmapsto [\,(\mathcal H\circ\iota^\bullet)(\alpha)\,],
\]
so no topological information is lost or gained by passing to the homothetic dressing—only the representatives within each cohomology class are transformed.

\subsubsection*{Homothetic de Rham Cohomology} 

Importantly, the nilpotency of $\hat{d}$ allows us to define a homothetic de Rham cohomology:
\begin{equation}\label{homCoh}
\hat{H}^k(M) := \ker(\hat{d}^k)/\text{Im}(\hat{d}^{k-1}),
\end{equation}
which encodes the topological information of the manifold as filtered through the homothetic structure. While the underlying topology of $M$ remains unchanged, the representatives of each cohomology class are now transformed by the dilaton field $\lambda(x)$ and the offset $\alpha_d$. For $\lambda = 0$, the ordinary de Rham theory is retrieved. However, for $\lambda \neq 0$, the cohomology classes are parametrized by homothetically equivalent forms.

This leads us to the homothetic Hodge decomposition:
\begin{equation}\label{Hom_Hodg_Dec}
\boxed{
\hat{\Omega}^k(M) = \text{Im}~\hat{d}^{k-1}~ \oplus~ \text{ker}~\hat{\Delta}^k~ \oplus ~\text{Im}~\hat{\delta}^{k+1},}
\end{equation}
in which the space of homothetically transformed $k$-forms splits orthogonally into exact, coexact, and harmonic components, all defined relative to the Weyl-rescaled metric $\tilde{\eta} = e^{-2\lambda(x)}\eta$. This decomposition underlies the forthcoming formulation of field equations, most notably, HMEs. 

\medskip

\noindent\textbf{Relation to prior work.}
Constructions closely related to ours appear in several literatures. 
In Hodge–de Rham theory, the Witten deformation considers the twisted differential
\(d_f := e^{-f}\, d\, e^{f} = d + df \wedge\) and its Hodge–Witten Laplacian, yielding orthogonal decompositions on weighted manifolds and inspiring a broad PDE/spectral theory for “drifted’’ Hodge Laplacians on differential forms \cite{Witten1982,BrandingWeighted2024}. 
From a cohomological viewpoint, Morse–Novikov (Lichnerowicz) cohomology twists by a closed one–form \(\theta\) via \(d_\theta := d + \theta \wedge\), extending the Witten picture from exact \(\theta=df\) to non-exact classes and linking topology with analysis of the associated Laplacians \cite{BurgheleaHaller2001,Otiman2016}. 
In the conformal geometry, there are conformally natural de Rham-type complexes and “conformal harmonic’’ theories (e.g. Branson–Gover and subsequent work) that situate Hodge theory within conformal/Weyl geometry \cite{BransonGover2005,Gover2004,AubryGuillarmou2008}. 
Finally, the functional-analytic framework for Hodge decompositions on (weighted) \(L^2\) spaces is cleanly captured by the Hilbert–complex framework \cite{BruningLesch1992,RodriguezHilbertComplexSurvey2023}. 
Together, these works cover twisted/weighted differentials, cohomology with local coefficients, conformal covariance, and abstract Hodge theory.

\medskip
\noindent\textbf{What is new here.}
Our approach is different in both formulation and emphasis. We work on a \emph{Weyl-integrable} background with Weyl one–form \(d\lambda\) and introduce a \emph{homothetic dressing} that embeds fields into pairs and acts through triangular \emph{block} operators which are \emph{conjugate} to direct-sum de Rham operators via \(\Lambda_w\) (so that, formally, \(\hat d\,\Lambda_w=\Lambda_w\,\overline d\)). This block–Weyl embedding enforces \emph{homothetic covariance} and introduces off-diagonal mixing that is absent from standard Witten/Morse–Novikov twists acting on a single copy of \(\Omega^\bullet\).
Analytically, on compact manifolds (and with the usual boundary conditions) our “homothetic” Hodge decomposition aligns with weighted Hodge theory— {\it e.g.}, using \(d_\lambda=e^{-\lambda} d e^{\lambda}\), \(\delta_\lambda=e^{\lambda}\delta e^{-\lambda}\)—but conceptually the \emph{Weyl-integrable, block-operator calculus} and its commuting-diagram isomorphism are, to the best of our knowledge, new and tailored to the HGT framework developed here (including the later homothetic gauge-theoretic couplings). 
Thus, while the cohomological content coincides with the de Rham/Morse–Novikov picture for \(\theta=d\lambda\), the \emph{homothetic, Weyl-covariant formulation and operator-level implementation} distinguish our contribution from existing treatments (see~\cite{Witten1982,BurgheleaHaller2001,BransonGover2005,BruningLesch1992}.)

\section{Homothetic Electromagnetism}\label{section:Homot_Elect}

The homothetic Hodge--de Rham theory developed in $\S$~\ref{section:4- Homothetic Hodge} provides a natural framework for constructing the homothetic $U(1)$ gauge theory, that is the homothetic electromagnetism; and in this regard, the chain isomorphism $\mathcal{H}\circ\iota^\bullet$ provides a guide in definition of the new action. For the reminder of this paper we use the normalized Weyl weight $w=1$ to obtain a more clear formalism, and we set $\Lambda:=\Lambda_{w=1}$. The general $w$ may be used for the calibration purposes of particular physical models.\\   

By definition of the projection $\pi_1:\hat{\Omega}\to\Omega$, and starting from the genuine Maxwell functional on $(M,\eta)$,
\begin{equation}
S_{\rm Max}[A;J]=-\frac{1}{2}\int F\wedge *_{\eta}F+\int A\wedge *_{\eta}J,
\qquad F:=dA,\quad d(*_{\eta}J)=0,
\label{eq:Max_bare}
\end{equation}
the HGT action ({\it i.e.}, the homothetically invariant action) is obtainable via pullback of $S_{\rm Max}$ along $\pi_1\circ\Lambda[\lambda]^{-1}$:
\begin{equation}
\
S_{\rm HGT}[\tilde A,A_d;\lambda]
=\bigl(\pi_1\circ\Lambda[\lambda]^{-1}\bigr)^{*}S_{\rm Max}
= S_{\rm Max}\!\left(A_d+e^{\lambda}(\tilde A-A_d);\,J\right).
\label{eq:SHGT_pullback}
\end{equation}
In $d=4$ the kinetic term is pointwise conformally invariant, so one may equivalently use $\tilde\eta=e^{-2\lambda}\eta$. With $\hat d$ and $\hat\delta$ as in~$\S$~\ref{sub-section:4.2}, the field-strength doublet is
\begin{equation}
\hat F:=\hat d\,\hat A
=\begin{pmatrix}\tilde F\\ F_d\end{pmatrix},
\qquad
\tilde F=d\tilde A + b\wedge(\tilde A-A_d),\quad F_d=dA_d,\quad b:=d\lambda.
\label{eq:blockF_def}
\end{equation}
Using the block pairing built from $*_{\tilde\eta}$ on each slot, the HGT Maxwell action reads
\begin{equation}
S_{\rm HGT}[\hat A;\lambda;\hat J]
= -\frac{1}{2}\int\!\big(\tilde F\wedge *_{\tilde\eta}\tilde F+F_d\wedge *_{\tilde\eta}F_d\big)
+\int\!\big(\tilde A\wedge *_{\tilde\eta}\tilde J + A_d\wedge *_{\tilde\eta}J_d\big),
\label{eq:HGT_action}
\end{equation}
with source doublet $\hat J=(\tilde J,J_d)^{\mathsf T}$ obeying $\hat\delta\,\hat J=0$ (upper slot couples to the physical current).

Varying \eqref{eq:HGT_action} and integrating by parts with $\hat\delta$ gives the homothetic Maxwell equations:
\begin{equation}
\boxed{\qquad
\hat\delta\,\hat F=\hat J,\qquad \hat d\,\hat F=0. \qquad}
\label{eq:block_Maxwell}
\end{equation}
In components:
\begin{align}
\delta_{\tilde\eta}\tilde F + \iota_{b^\sharp}\tilde F &= \tilde J, &
d\tilde F + b\wedge\tilde F - b\wedge F_d &= 0, \label{eq:upper_eqs}\\
\delta_{\tilde\eta}F_d - \iota_{b^\sharp}\tilde F &= J_d, &
dF_d&=0. \label{eq:lower_eqs}
\end{align}
Setting $J=0$ (equivalently $\hat J=0$) in \eqref{eq:block_Maxwell} yields the homogeneous system
\begin{equation}
\hat\delta\,\hat F=0,\qquad \hat d\,\hat F=0,
\label{eq:Homog_block}
\end{equation}
that is,
\begin{equation}
\delta_{\tilde\eta}\tilde F + \iota_{b^\sharp}\tilde F = 0,\qquad
d\tilde F + b\wedge\tilde F - b\wedge F_d = 0,\qquad
\delta_{\tilde\eta}F_d - \iota_{b^\sharp}\tilde F = 0,\qquad
dF_d=0.
\label{eq:Homog_components}
\end{equation}

\begin{remark}[Conformal covariance]
In four spacetime dimensions, the homogeneous homothetic Maxwell system \eqref{eq:Homog_block} or \eqref{eq:Homog_components} is conformally invariant—precisely in the same sense as the classical source-free Maxwell equations.
\end{remark}

\textbf{Homothetic Stress-Energy Tensor.}
Varying the full HGT action \eqref{eq:HGT_action} with respect to $\tilde\eta^{\mu\nu}$ (using the block pairing from $*_{\tilde\eta}$) gives
\[
\hat T_{\mu\nu}
=\tfrac14\tilde\eta_{\mu\nu}\!\left(\tilde F_{\alpha\beta}\tilde F^{\alpha\beta}
+F_{d,\alpha\beta}F_d^{\alpha\beta}\right)
-\big(\tilde F_{\mu\alpha}\tilde F_{\nu}{}^{\alpha}
+F_{d,\mu\alpha}F_{d,\nu}{}^{\alpha}\big),
\]
with indices raised by $\tilde\eta$. Using \eqref{eq:block_Maxwell} and the homogeneous identities ($d\tilde F+b\wedge\tilde F-b\wedge F_d=0$, $dF_d=0$), its divergence is
\[
\tilde\nabla^\mu \hat T_{\mu\nu}
= -\,\tilde F_{\nu\alpha}\,\tilde J^{\alpha}
-\,F_{d,\nu\alpha}\,J_d^{\alpha}
\;+\;\underbrace{\big(\tilde F_{\nu\alpha}-F_{d,\nu\alpha}\big)\,\big(\iota_{b^\sharp}\tilde F\big)^{\alpha}}_{\;0\ \text{by }\eqref{eq:upper_eqs}--\eqref{eq:lower_eqs}}.
\]
Hence, in vacuum $(\hat J=0)$ we have $\tilde\nabla^\mu \hat T_{\mu\nu}=0$.

\textbf{Three-vector source-free formulation.}
For our next usage, and because it is more convenient, we obtain as well, the conventional three-vector formalism of Maxwell equations. To this end, we choose global inertial coordinates $(t,\mathbf{x})$ on $(M,\eta)$ and decompose
\[
\tilde A_\mu=(\tilde\phi,\tilde A),\qquad A_{d,\mu}=(\phi_d,A_d),\qquad
b_\mu=(\partial_t\lambda,\nabla\lambda).
\]
(Spatial vectors are written without arrows.) Then, by definition of the generalized fields
\begin{equation}
\begin{aligned}
\tilde E &:= -\partial_t \tilde A - \nabla\tilde\phi\;-\;(\partial_t\lambda)\,(\tilde A-A_d)\;+\;(\tilde\phi-\phi_d)\,\nabla\lambda,\\[2pt]
\tilde B &:= \nabla\times \tilde A\;+\;(\nabla\lambda)\times(\tilde A-A_d),
\end{aligned}
\label{eq:EB_defs}
\end{equation}
and, for the reference slot, $E_d:=-\partial_t A_d-\nabla\phi_d$, $B_d:=\nabla\times A_d$,
the homogeneous generalized Maxwell equations \eqref{eq:Homog_components} translate to
\begin{empheq}[left=\empheqlbrace]{align}
\nabla\cdot \tilde E\;+\;(\nabla\lambda)\cdot(\tilde E - E_d) &= 0,
\label{eq:GM_Gauss}\\[4pt]
\partial_t \tilde E\;-\;\nabla\times \tilde B\;+\;(\partial_t\lambda)\,(\tilde E - E_d)\;-\;(\nabla\lambda)\times(\tilde B - B_d) &= 0,
\label{eq:GM_Ampere}\\[4pt]
\partial_t \tilde B\;+\;\nabla\times \tilde E\;+\;(\partial_t\lambda)\,(\tilde B - B_d)\;+\;(\nabla\lambda)\times(\tilde E - E_d) &= 0,
\label{eq:GM_Faraday}\\[4pt]
\nabla\cdot \tilde B\;+\;(\nabla\lambda)\cdot(\tilde B - B_d) &= 0,
\label{eq:GM_GaussMag}
\end{empheq}
which reduce to the standard sourceless Maxwell equations when $\lambda$ is constant (so $b=d\lambda=0$) and $A_d$ is fixed (hence $E_d=B_d=0$).\\

\textbf{Remarks.}
\begin{itemize}
\item[(i)] \textbf{Gauge symmetry:} for any scalar doublet $\hat\chi=(\chi,\chi_d)$, the transformation $\hat A\mapsto \hat A+\hat d\,\hat\chi$ preserves $\hat F$ and $S_{\rm HGT}$, inducing the usual shift $\tilde A\mapsto \tilde A+d\chi$ in the upper slot. 
\item[(ii)] \textbf{Reduction:} if $\lambda$ is constant, \eqref{eq:GM_Gauss}–\eqref{eq:GM_GaussMag} split into two decoupled Maxwell systems for $(\tilde E,\tilde B)$ and $(E_d,B_d)$; fixing $A_d$ collapses the upper slot to the genuine sourceless Maxwell theory on $(M,\eta)$.
\item[(iii)] \textbf{Computational EM:} system \eqref{eq:GM_Gauss}--\eqref{eq:GM_GaussMag} provides a computational framework with the potential of using in the computational electrodynamics. A brief investigation of its mathematical properties and its relation with the penalized Maxwell equations are provided in \ref{HME-Penal}. 
\end{itemize} 

\medskip

\textbf{Related literature.}
Two main directions are observable in the literature that come closest to the present homothetic formulation. First, a large body of work has pursued modified electrodynamics. Classic examples are the Born–Infeld (BI) nonlinearity, which saturates the electric field and renders the point-charge self-energy finite \cite{BornInfeld1934,SorokinReview2022}, generalized ModMax–type theories with conformal/duality symmetries that can also regularize the field in certain parameter ranges \cite{KruglovModMax21}, and higher-derivative but linear Bopp–Podolsky/BLTP electrodynamics, which yields a finite self-energy and a Yukawa-regularized Coulomb law \cite{Bopp1940,podolsky1942}. In curved spacetime, nonlinear electrodynamics has been used as matter to build regular (non-singular) charged geometries \cite{AyonBeatoGarcia1998,BronnikovReview2022}. It should be emphasized that although by embedding in the double space we ultimately restore the linearity, however, HGT still classifies as a non-linear (homothetic) electrodynamics theory.

Second, couplings between a scalar (``dilaton'') and the electromagnetic field are well–studied. In 4D effective actions descending from string/Kaluza–Klein reductions, one generically finds a non-minimal gauge kinetic function, typically of the form $e^{-2a\phi} F_{\mu\nu}F^{\mu\nu}$, leading to the Einstein–Maxwell–dilaton (EMD) family of models and their black–hole solutions \cite{GibbonsMaeda1988,GarfinkleHorowitzStrominger1991}. Broader axion–dilaton couplings and duality structures were developed in \cite{GibbonsRasheed1995}, while varying–$\alpha$ frameworks in which a scalar controls an effective permittivity/permeability go back to Bekenstein \cite{Bekenstein1982}. 

There are also recent studies of electromagnetism in Weyl(-integrable) geometries and Weyl-invariant frameworks, but these typically couple the gauge fields to an independent Weyl gauge field rather than to a homothetic dressing driven by a scalar gradient \cite{Mavrogiannis2024,Oancea2024}. 

\subsection{The Offset fields}\label{sec:offset}

A key feature of HGT is the introduction of offset fields, such as $A_d$ in the homothetic $U(1)$ theory. By definition, $A_d$ acts as the ``homothety center'' for the affine dressing \eqref{eq:HDD_1}, that is, the reference potential from which the physical field $\tilde{A}$ is measured. However, unlike fixed background fields or pure gauge parameters, the offset fields in HGT have their own dynamics, which is imposed by embedding into the doubled space $\bar{\Omega}^\bullet(M)$.

As seen in the action \eqref{eq:HGT_action}, $A_d$ possesses its own kinetic term ($F_d \wedge \star F_d$) and must satisfy the homothetic Maxwell equations. Specifically, the lower-slot equation of motion,
\[
\delta_{\tilde\eta}F_d - \iota_{b^\sharp}\tilde F = J_d, \qquad dF_d=0,
\]
reveals that the offset field is not only propagating but is also sourced by the physical field $\tilde{F}$ (mediated by the dilaton gradient $b=d\lambda$). Consequently, the fundamental object of the theory is neither $\tilde{A}$ nor $A_d$ in isolation, but the coupled gauge doublet $\hat{A}$. The offset field is thus an irreducible, dynamical component of the homothetic gauge redundancy, which, as shown in $\S$\ref{section:8}, provides the necessary degrees of freedom to regularize singular sources.

\section{The Homothetic Lorenz gauge and Potential}\label{section: Homot-Cohomo}

We now turn to the dynamics of the homothetic gauge fields, where some interaction terms appear canonically. As one can see, these are the very penalty terms used to impose boundary conditions on the Laplace and wave equations. We shall discuss the homogeneous cases, but extension to the non-homogeneous case is almost immediate.

\subsection{Homothetic Lorenz Gauge}

Starting from the homothetic Maxwell equations (\ref{eq:Homog_block}), and using \( \hat{F} = \hat{d} \hat{A} \), we define and choose the ``homothetic'' Lorenz gauge
\begin{equation}\label{eq:Lorentz_gauge}
\boxed{\hat{\delta} \hat{A} = 0,}
\end{equation}
as our ``gauge fixing'', in which $\hat{A}:=( \tilde{A}~~A_d)^{\mathsf{T}}$, as we defined earlier. Then, the ``homothetic'' residual gauge freedom is any block scalar potential $\hat{\phi}:=( \tilde{\phi}~~\phi_d)^{\mathsf{T}}$ satisfying   
\begin{equation}\label{eq:Hom_pot}
\boxed{
\hat{\delta} \; \hat{d} \; \hat{\phi}=0.}
\end{equation}
The dynamics of the ``homothetic gauge potentials'' obey this equation.

\subsection{Homothetic Harmonic Gauge Fields and Potentials}\label{sub-section:6.1}

In particular, the dynamics of the physical components of the homothetic variables (the first row of the homothetic variables) are desired. To this end, from \eqref{eq:Lorentz_gauge} one obtains:
\[
\delta \tilde{A} - * (d\lambda) \wedge * (\tilde{A} - A_d) = 0.
\]
On the Minkowski spacetime, this yields a generalized conservation-like equation:
\begin{equation}\label{eq:Conservation}
\boxed{
\nabla \cdot \tilde{A} - \nabla \lambda \cdot (\tilde{A} - A_d) = 0.}
\end{equation}
In spacetime index notation:
\[
\partial_\mu \tilde{A}^\mu - \partial_\mu \lambda (\tilde{A}^\mu - A_d^\mu) = 0,
\]
which modifies the classical condition \( \partial_\mu A^\mu = 0 \) by a geometric source term related to \( \lambda \).

In the same way, one can derive the homothetic wave and Laplace equations for \( \tilde{\phi} \), from the first row of \eqref{eq:Hom_pot}. By definition of
\[
\square := \partial_\mu \partial^\mu = -\partial_t^2 + \nabla^2, \quad
|\partial \lambda|^2 := \partial_\mu \lambda \, \partial^\mu \lambda, \quad
(\partial \lambda \cdot \partial f) := \partial_\mu \lambda \, \partial^\mu f.
\]

\subsubsection*{For the general non-stationary fields}

\[
\boxed{
\square \tilde{\phi}
+ 2\,\partial_\mu \lambda (\partial^\mu \tilde{\phi} - \partial^\mu \phi_d)
+ (\square \lambda + |\partial \lambda|^2) (\tilde{\phi} - \phi_d) = 0.
}
\]

\subsubsection*{For the stationary fields}

Assuming \( \lambda \) and the potentials are time-independent:
\begin{equation}\label{eq:modif_Lap_eq}
\boxed{
\nabla^2 \tilde{\phi}
+ 2\,\nabla \lambda \cdot (\nabla \tilde{\phi} - \nabla \phi_d)
+ (\nabla^2 \lambda + |\nabla \lambda|^2) (\tilde{\phi} - \phi_d) = 0.
}
\end{equation}
As one can see, these equations exhibit new coupling between the scalar potential components \( \tilde{\phi} \) and \( \phi_d \), mediated by \( \lambda \).

As one can see, the modified wave and Laplace equations derived above contain the additional interaction terms. Assuming there is a boundary \(\partial\Omega\), if \(\lambda\) is chosen such that locally on the boundary we have 
\begin{equation}\label{eq:penal_wave_1}
\nabla \lambda = \beta\,\mathbf{n}\quad\text{with }\ \beta\to\infty,
\end{equation}
while \(\nabla\lambda \to 0\) in the interior, then these terms take the role of the penalizing terms on the boundary enforcing simultaneously
\begin{equation}\label{eq:penal_wave_2}
\tilde{u}\big|_{\partial\Omega}=u_d\big|_{\partial\Omega},
\qquad
\partial_{\mathbf{n}}\tilde{u}\big|_{\partial\Omega}=\partial_{\mathbf{n}}u_d\big|_{\partial\Omega},
\end{equation}
that is, they act as boundary penalization terms that impose Cauchy boundary conditions on both the wave and Laplace equations (in the sense of penalty/SAT/weak imposition methods). This connects the HGT, with such a choice, with classical penalty and Nitsche/SAT formulations for enforcing boundary conditions. See, e.g,  \cite{Nitsche1971,CarpenterGottliebAbarbanel1994,BrennerScott2007,GlowinskiLionsTremolieres1981} for background.

\section{Application: A Non-Singular Point Electric Charge}\label{section:8}

Classical electrodynamics has long faced a well-known difficulty: the infinite energy of the vacuum associated with point charges~\cite{rohrlich1990,frisch2008}. It has been a foundational issue in classical field theory, prompting numerous attempts at regularization, ranging from
different classes of non-linear electrodynamics (NED)~\cite{BornInfeld1934,Kruglov2014}, addition of the higher derivatives~\cite{Bopp1940,podolsky1942}, inclusion of non-local effects via fractional derivatives~\cite{tarasov2022,Efimov1972} (see \cite{Kruglov2024} for a brief and good survey), to the models that ascribe a finite size to the electron offer a mechanical resolution to the self-energy divergence~\cite{rohrlich1990,spohn2004,Dirac1962}. In quantum electrodynamics (QED), the divergence is softened through vacuum polarization effects. The effective Heisenberg–Euler~\cite{Heisenberg1936} and Schwinger~\cite{schwinger1951} Lagrangians describe nonlinear photon-photon interactions that act similarly to NED under strong-field conditions. Incorporating gravity provides yet another route. Coupling NED models to general relativity has shown that both the field and spacetime singularities may be smoothed~\cite{Hendi2012,Kruglov2016-prd, Kruglov2015}. These \emph{gravity-coupled models} often yield black hole solutions with regular centers. In the framework developed in the previous section, a new perspective emerges.

Within HGT, the field equations naturally admit boundary conditions (see \eqref{eq:penal_wave_1}--\eqref{eq:penal_wave_2}, and \ref{HME-Penal}). This allows us to model a point charge not as a singular forcing term \(\rho=Q\,\delta_0\), but instead via a \emph{boundary law} on a small sphere \(S_R=\partial B_R(0)\) and then pass to the limit \(R\to0\). This “small-sphere” construction is topologically supported (all data are posed on an \(S^2\)), and is equivalent to the operator-theoretic viewpoint of \emph{self-adjoint extensions} (SAE), in which short-distance physics is encoded by a boundary relation rather than by an explicit \(\delta\)-source (see, {\it e.g.}, ~\cite{AlbeverioPoint,GitmanTyutinVoronov2012}). We work in the static, spherically symmetric case. 

Let \(\tilde{\mathbf E}(r)=\tilde E_r(r)\,\hat{\mathbf r}\) be the HGT electric field, \(\lambda=\lambda(r)\) the homothetic (dilaton) field (allowed to have a jump at the origin), and \(\mathbf E_d(r)=E_{d,r}(r)\,\hat{\mathbf r}\) a bounded reference field with \(E_{d0}:=\lim_{r\to0}E_{d,r}(r)\) finite. The homothetic Gauss law can be written in weighted form
\begin{equation}
\nabla\!\cdot\!\big(e^{\lambda}\tilde{\mathbf E}\big)
= e^{\lambda}\rho \;+\; e^{\lambda}(\nabla\lambda)\!\cdot\!\mathbf E_d.
\label{eq:weighted-gauss}
\end{equation}
We model the point source as a ball \(B_R(0)\) and imposing a boundary law on \(S_R=\partial B_R\), then send \(R\to0\).

Integrating \eqref{eq:weighted-gauss} over \(B_R(0)\) and using the divergence theorem gives
\begin{equation}
\oint_{S_R} e^{\lambda}\tilde{\mathbf E}\!\cdot\!\hat{\mathbf n}\,dS
=\int_{B_R} e^{\lambda}\rho\,dV + \int_{B_R} e^{\lambda}(\nabla\lambda)\!\cdot\!\mathbf E_d\,dV.
\end{equation}
Allow a point jump in \(\lambda\) at \(r=0\):
\(
\lambda'(r)=\kappa\,\delta(r)
\)
so that \(\int_{B_R} e^{\lambda}(\nabla\lambda)\!\cdot\!\mathbf E_d\,dV
= e^{\lambda(0)}\,\kappa\,E_{d0}\).
For a point charge \(Q\), \(\int_{B_R} e^{\lambda}\rho\,dV= e^{\lambda(0)}Q\). Hence, for every \(R>0\),
\begin{equation}
4\pi R^2\, e^{\lambda(R)} \,\tilde E_r(R^+) \;=\; e^{\lambda(0)}\!\left(Q + \kappa\,E_{d0}\right)
\;=:\; Q_{\mathrm{eff}}.
\label{eq:BC-small-sphere}
\end{equation}
Equation \eqref{eq:BC-small-sphere} is the \emph{boundary condition} on \(S_R\): the weighted flux equals the effective enclosed charge (the explicit \(Q\) plus the contribution from the \(\lambda\)-jump through \(E_{d0}\)). This is the small-sphere version of fixing the SAE parameter \cite{AlbeverioPoint}.

\textbf{Exterior equation and solution for \(r>R\).}
Outside the core there is no volumetric source, so \eqref{eq:weighted-gauss} reduces to
\begin{equation}
\frac{1}{r^{2}}\frac{d}{dr}\!\Big(r^{2} e^{\lambda(r)} \tilde E_r(r)\Big)
= e^{\lambda(r)}\lambda'(r)\,E_{d,r}(r), \qquad r>R.
\label{eq:radial-weighted}
\end{equation}
If \(\lambda'\) is supported only at the origin (pure jump at \(r=0\)), then the right-hand side of \eqref{eq:radial-weighted} vanishes for \(r>R\) and
\begin{equation}
r^{2} e^{\lambda(r)} \tilde E_r(r) \equiv \frac{Q_{\mathrm{eff}}}{4\pi} \quad\Longrightarrow\quad
\boxed{\;\tilde E_r(r)= \dfrac{Q_{\mathrm{eff}}}{4\pi}\,\frac{e^{-\lambda(r)}}{r^{2}},\quad r>R.\;}
\label{eq:field-solution}
\end{equation}
(If \(\lambda'\) has compact support in a small neighborhood of the origin, the same formula holds for \(r\) beyond that support, and one matches it to \eqref{eq:BC-small-sphere}.)

\textbf{Finite self-energy.}
In units where the electrostatic energy density is \(\tfrac12|\tilde{\mathbf E}|^{2}\), the self-energy outside the core is
\begin{equation}
U(R) = \frac{1}{2}\int_{r>R} |\tilde{\mathbf E}|^{2}\,d^{3}x
= \frac{Q_{\mathrm{eff}}^{2}}{8\pi}\int_{R}^{\infty}\frac{e^{-2\lambda(r)}}{r^{2}}\,dr.
\label{eq:self-energy}
\end{equation}
Assume near the origin
\(
\lambda(r) \sim -\alpha \ln r \quad (r\to0^+)
\),
so \(e^{-2\lambda(r)}\sim r^{2\alpha}\). Then the integrand behaves like \(r^{2\alpha-2}\), and the core contribution \(\int_{0}^{\epsilon} r^{2\alpha-2}dr\) converges iff
\begin{equation}
\boxed{\;\alpha>\frac{1}{2}\;}\quad\text{(finite self-energy).}
\label{eq:alpha-threshold}
\end{equation}
Moreover, \(\tilde E_r(r)\sim r^{\alpha-2}\) from \eqref{eq:field-solution}, so the field itself remains finite at \(r=0\) provided \(\alpha\ge 2\).

\textbf{Example (finite field and energy).}
Choose a core scale \(r_c>0\) and set
\begin{equation}
\lambda(r)=
\begin{cases}
-2\ln\!\big(r/r_c\big), & 0<r\le r_c,\\
\text{smoothly transition to }0, & r\ge r_c,
\end{cases}
\qquad\Rightarrow\qquad
\tilde E_r(r)\xrightarrow[r\to0]{} \dfrac{Q_{\mathrm{eff}}}{4\pi r_c^{2}}.
\end{equation}
Then \eqref{eq:self-energy} gives a finite total:
\begin{equation}
\boxed{\;U(0^+)=\dfrac{Q_{\mathrm{eff}}^{2}}{6\pi\,r_c}\;<\infty.\;}
\end{equation}
This realizes a non-singular “point” charge via honest boundary data on \(S_R\) for each \(R>0\), with the point limit \(R\to0\) taken at the end. In the SAE language, \eqref{eq:BC-small-sphere} fixes the extension parameter (the weighted flux at the origin), and the exterior PDE carries no explicit \(\delta\)-source.


\section{Conclusions and Outlook}\label{conclusions}

In this work, we have introduced a Homothetic Gauge Theory (HGT), built upon a novel homothetic dilaton dressing of the fields on the Weyl integrable geometry, formulated for a general Weyl weight $w$. By embedding the dressed fields into a doubled space, we were able to linearize the affine homothetic dressing. This construction allowed for the development of a corresponding homothetic Hodge--de Rham theory. Definition of a set of covariant block operators, covariant with respect to the homothetic transformation, leaded to a homothetic Hodge decomposition. While the homothetic de Rham complex is topologically isomorphic to the ordinary de Rham complex, its harmonic representatives are fundamentally different, as they are dressed by the homothetic dilaton field $\lambda$ and governed by the chosen weight.

The primary application of this formalism, obtained by specializing to the weight $w=1$, was the construction of a homothetic electromagnetism. The resulting homothetic Maxwell equations (HMEs) describe a coupled system for the physical gauge field and the offset field, which we have shown to be a dynamical component of the homothetic gauge freedom. We demonstrated that this theory provides a compelling classical solution to the long-standing point-charge self-energy problem. By modeling the point charge as a boundary condition on an infinitesimal sphere---by employing the theory of self-adjoint extensions---we showed that a suitable choice for the dilaton profile ($\lambda(r) \sim -\alpha \ln r$ with $\alpha > 1/2$) renders both the electric field and its total self-energy finite at the origin.

Furthermore, we highlighted a remarkable connection between the HGT framework and computational methods. The interaction terms that emerge naturally from the homothetic Lorenz gauge are mathematically identical to the penalty terms used in numerical methods (such as Nitsche's method or SATs) to weakly enforce boundary conditions. This suggests that the homothetic dilaton field $\lambda$ can be interpreted as a physical penalization field that regularizes the theory.

This framework opens several intriguing avenues for future research.
\begin{itemize}
    \item \textbf{Quantization:} The most immediate direction is the quantization of HGT. Investigating the quantum properties of the doubled-field system, the quantum-field-theoretic role of the offset field as a dynamical gauge component, and whether the classical regularization of the point charge persists at the quantum level are all critical questions.

    \item \textbf{Non-Abelian Theories:} The formalism presented here was limited to a $U(1)$ gauge theory. Its extension to non-abelian theories, such as a Homothetic Yang-Mills theory, would be a significant development, potentially offering new insights into the behavior of non-abelian fields and their singularities.

    \item \textbf{Coupling to Gravity:} The theory is intrinsically geometric, being formulated on a Weyl-integrable spacetime. A natural next step is to couple HGT to a dynamical gravitational field. Promoting $\lambda$ from a background dilaton to a dynamical scalar field could have profound implications, particularly in the context of Einstein-Maxwell-Dilaton theories, black hole physics, and the resolution of spacetime singularities.

    \item \textbf{Cosmological Implications:} The presence of a fundamental scalar field $\lambda$ invites cosmological exploration. Investigating whether the homothetic dilaton could play a role in dark energy (as a quintessence field) or in the early universe (as an inflaton) is a promising direction.

    \item \textbf{Computational Physics:} The explicit link between the HMEs and numerical penalty methods warrants deeper study. It is possible that HGT could provide a physical principle for designing new, robust, and physically-motivated boundary conditions or stabilization techniques for computational electromagnetism.
\end{itemize}

In summary, Homothetic Gauge Theory offers a mathematically consistent and physically motivated framework that unifies field regularization with a geometric dressing, providing new tools to address classical problems and opening new pathways for theoretical and computational exploration.

\appendix

\section{Principal \texorpdfstring{$\mathcal{H}$}{H}-bundle formulation and curvature}
\label{app:principal_H_bundle}

Let $\mathcal{H}\cong ({\mathbb R}_{>0},\cdot)$ be the \emph{homothetic group} acting  
on each doubled fibre by block-diagonal multiplication.%
\footnote{Locally---on a chart $(U,x)$---a homothety is specified by a smooth
function $\lambda\in C^\infty(U)$, the action being $v\mapsto e^{\lambda(x)}v$. 
Pointwise multiplication turns $\mathcal{H}$ into a $1$-dimensional
abelian Lie group with Lie algebra $\mathfrak h\cong {\mathbb R}$ under addition.}
Choose a good open cover $\{U_\alpha\}_{\alpha\in A}$ of $M$ and define
transition functions
\[
   g_{\alpha\beta}\colon U_\alpha\cap U_\beta \;\longrightarrow\; \mathcal{H},
   \qquad
   g_{\alpha\beta}(x)\;=\;e^{\lambda_{\alpha\beta}(x)}, 
   \quad
   \lambda_{\alpha\beta}\in C^\infty(U_\alpha\cap U_\beta),
\]
satisfying the usual cocycle conditions
$g_{\alpha\alpha}=1,\;g_{\alpha\beta}g_{\beta\gamma}=g_{\alpha\gamma}$.
They define a principal $\mathcal{H}$-bundle\,
$\pi\colon P\to M$ with local trivializations
$\psi_\alpha:\pi^{-1}(U_\alpha)\!\to\!U_\alpha\times\mathcal{H}$
and right action $(p,h)\!\mapsto\!ph$.
Throughout we fix a \emph{principal connection}
\[
   \omega\;\in\;\Omega^{1}(P,\mathfrak h),
   \qquad
   R\,=\,d\omega\;\in\;\Omega^{2}(P,\mathfrak h),
\]
characterised by:
\[
   \text{\emph{(i)}}\;\;\;R_{X^\#}\omega=1
   \quad\text{and}\quad
   \text{\emph{(ii)}}\;\;\;R_{h}^{*}\omega=Ad_{h^{-1}}\!\omega
   \;(=\omega\text{ since }\mathcal{H}\text{ is abelian}),
\]
with $X^\#$ the fundamental field generated by $X\in\mathfrak h$.

\textbf{Local expression.}
Choosing local sections $\sigma_\alpha\colon U_\alpha\to P$ we obtain the 
connection $1$-forms
\[
   \Gamma_\alpha\;:=\;\sigma_\alpha^{\!*}\omega\;\in\;\Omega^{1}(U_\alpha,\mathfrak h)
   \quad\text{and}\quad
   R_\alpha\;:=\;\sigma_\alpha^{\!*}R\;=\;d\Gamma_\alpha.
\]
Under a change of trivialization one has
\begin{equation}
   \Gamma_\beta
   \;=\;
   g_{\alpha\beta}^{-1}\Gamma_\alpha\,g_{\alpha\beta}
   +g_{\alpha\beta}^{-1}\,dg_{\alpha\beta}
   \;=\;
   \Gamma_\alpha + d\lambda_{\alpha\beta},\label{eq:Gamma_gauge}
\end{equation}
because $Ad_{g_{\alpha\beta}^{-1}}$ is trivial.  Curvature transforms
accordingly as $R_\beta=R_\alpha$, so $R$ is \emph{gauge-invariant}.
If we write $\Gamma_\alpha=d\lambda_\alpha$ for a smooth scalar
$\lambda_\alpha\in C^\infty(U_\alpha)$, then \eqref{eq:Gamma_gauge} is
consistent with $\lambda_\beta=\lambda_\alpha+\lambda_{\alpha\beta}$.
Hence $R=d^{2}\lambda_\alpha=0$ locally, and globally
\[
   R\;=\;d\Gamma + \Gamma\wedge\Gamma \quad
   (\text{Maurer–Cartan form}),
\]
establishing Eq.~(3.6) in a manifestly \emph{principal-bundle}
language.  The global existence of $\Gamma$ is equivalent to the
exactness of the Čech $1$-cocycle $\{\lambda_{\alpha\beta}\}$ in
$H^{1}(M,R)$.

\section{Computational Electromagnetism via the HMEs}\label{HME-Penal}

In this appendix, we demonstrate the correspondence between different standard boundary conditions of Maxwell equations and the penalty terms in the HMEs. We assume that the offset fields $F_d$ are given.

Consider the 3–vector formulation of the homothetic Maxwell system \eqref{eq:GM_Gauss}–\eqref{eq:GM_GaussMag} for the unknowns $(\tilde{\mathbf E},\tilde{\mathbf B})$, with the scalar $\lambda$ and the offset fields $(\mathbf E_d,\mathbf B_d)$ given (for simplicity, time–independent). Let $D\subset\mathbb{R}^3$ be a bounded domain with smooth boundary $\partial D$.

Assume that in a thin layer adjacent to $\partial D$ we have
\[
\nabla\lambda=\beta\,\mathbf n,\qquad \beta=\beta(x)\gg 1\quad\text{(large near $\partial D$),}
\]
with $\mathbf n$ the outward unit normal. Then the terms $(\nabla\lambda)\!\cdot(\tilde{\mathbf E}-\mathbf E_d)$, $(\nabla\lambda)\!\times(\tilde{\mathbf E}-\mathbf E_d)$ and their magnetic analogues play the role of ``penalty'' terms imposing the boundary conditions as $\beta$ increases. See for example, spectral/Fourier penalty and Nitsche/penalty methods in computational EM \cite{HesthavenPenalty2000,GalaguszShirokoffNave2016,BoffiCodinaTurk2025}.

\textbf{1- Finite–penalty regime ($\beta<\infty$)-- stability and dissipation.}
The formal energy estimate of \eqref{eq:GM_Ampere}–\eqref{eq:GM_Faraday} for $\tilde{\mathbf E}$ and $\tilde{\mathbf B}$, and definition of energy $\mathcal E(t):=\tfrac12\!\int_D (|\tilde{\mathbf E}|^2+|\tilde{\mathbf B}|^2)\,dx$, reads:
\[
\frac{d}{dt}\mathcal E(t)\;\le\;
-\,\int_{\partial D}\!\beta\Big(
|\mathbf n\!\times(\tilde{\mathbf E}-\mathbf E_d)|^2+|\mathbf n\!\cdot(\tilde{\mathbf E}-\mathbf E_d)|^2
+|\mathbf n\!\times(\tilde{\mathbf B}-\mathbf B_d)|^2+|\mathbf n\!\cdot(\tilde{\mathbf B}-\mathbf B_d)|^2
\Big)\,dS,
\]
up to usual interior sources. Thus, for each fixed $\beta > 0$, the penalized system is (formally) well–posed and ``dissipative'' ($\frac{d}{dt}\mathcal E(t)\leq 0$ because $\beta > 0$ and the terms in the integral are all non-negative sums of squares). It relaxes $(\tilde{\mathbf E},\tilde{\mathbf B})$ toward the prescribed offsets on $\partial D$. In discretizations, it is natural to choose $\beta$ so that the thin layer thickness is $\delta\sim\beta^{-1}=O(h)$ (as it has done for the Navier-Stokes equations \cite{Sabetghadam2015,Badri2020}).

\textbf{2- Infinite–penalty limit ($\beta\to\infty$): Cauchy data.}
In the limit $\beta\to\infty$, the penalty terms impose
\[
\mathbf n\!\times(\tilde{\mathbf E}-\mathbf E_d)=\mathbf 0,\quad
\mathbf n\!\cdot(\tilde{\mathbf E}-\mathbf E_d)=0,\quad
\mathbf n\!\times(\tilde{\mathbf B}-\mathbf B_d)=\mathbf 0,\quad
\mathbf n\!\cdot(\tilde{\mathbf B}-\mathbf B_d)=0 \quad\text{on }\partial D.
\]

This is, in fact, a representation of the Cauchy data (over–determined) on the boundary, for the first–order Maxwell IBVP \footnote{We should emphasize that we have a pre-assumption that $(\mathbf E_d,\mathbf B_d)$ are compatible. Here it means that they are coming from a target impedance law {\it or} from an exterior solution satisfying the jump/radiation conditions.}. Consequently, the penalized solution converges to the Maxwell solution with that boundary condition (cf.\ impedance and radiation conditions \cite{Senior1995,KirchnerUrbanZeeb2016}).
On the other hand, without altering the structure of the equations, the offset fields determine which conventional type of boundary conditions of Maxwell equations to be imposed:

\begin{enumerate}
\item \textbf{Impedance (Leontovich/Silver–Müller context).} Choose $(\mathbf E_d,\mathbf B_d)$ that themselves satisfy a surface law, e.g.\ Leontovich \( \mathbf n\times \mathbf H_d = Y_s\ \mathbf n\times(\mathbf n\times \mathbf E_d)\), or are consistent with Silver–Müller at a truncation boundary; then the penalty drives $(\tilde{\mathbf E},\tilde{\mathbf B})$ to that trace \cite{Senior1995,KirschHettlichNotes}.
\item \textbf{Full Field Dirichlet Conditions.} This framework enforces the full vector trace (Cauchy data), which is stronger than standard tangential boundary conditions.
\begin{itemize}
\item \textbf{Electric Dirichlet ($\tilde{\mathbf E}|_{\partial D} = \mathbf 0$):} Setting $\mathbf E_d \equiv \mathbf 0$ imposes the condition $\tilde{\mathbf E} = \mathbf 0$ on $\partial D$ (both tangential and normal components). This is distinct from a standard PEC condition, which only requires $\mathbf n \times \tilde{\mathbf E} = \mathbf 0$ \cite{JinFEM2014,Monk2003}.
\item \textbf{Magnetic Dirichlet ($\tilde{\mathbf B}|_{\partial D} = \mathbf 0$):} Setting $\mathbf B_d \equiv \mathbf 0$ imposes the condition $\tilde{\mathbf B} = \mathbf 0$ on $\partial D$. This is distinct from a standard PMC condition, which only requires $\mathbf n \times \tilde{\mathbf H} = \mathbf 0$ (or $\mathbf n \times \tilde{\mathbf B} = \mathbf 0$) \cite{JinFEM2014,Monk2003}.
\end{itemize}
\item \textbf{Absorbing/Truncation via Absorbing Boundary Condition(ABC)/Perfectly Matched Layer(PML).} Classical ABCs (Engquist–Majda) and PMLs (Bérenger) provide nonreflecting truncations without prescribing both field traces; here, by contrast, the thin-layer penalty is a Dirichlet clamp for the chosen offsets \cite{EngquistMajda1977,Berenger1994,BecacheJoly2002,JohnsonPML2008}.
\end{enumerate}


\end{document}